\documentclass[aps,prl, twocolumn,nofootinbib,10pt]{revtex4}
    \usepackage{graphicx}
    \usepackage{amsmath,amssymb,mathrsfs}
\usepackage{epsf,bm}
\usepackage{slashed}
\usepackage{epsfig}
\usepackage[font=small,skip=0pt,justification=raggedright]{caption}

\newcounter{fig}   \newcommand{\lbfig}[1]{\refstepcounter{fig}
\label{#1} }

\newcommand{\bea}{\begin{eqnarray}}
\newcommand{\eea}{\end{eqnarray}}
\newcommand{\be}{\begin{equation}}
\newcommand{\ee}{\end{equation}}

\def\bfph{{\pmb{\phi}}}

\newcommand{\re}[1]{(\ref{#1})}

\begin{document}

\title{Magnetic Skyrmions coupled to fermions}

\author{I.~Perapechka}
\affiliation{Department of Theoretical Physics and Astrophysics,
Belarusian State University, Minsk 220004, Belarus}
\author{Ya.~Shnir}
\affiliation{BLTP, JINR, Dubna 141980, Moscow Region, Russia\\
Department of Theoretical Physics, Tomsk State Pedagogical University, Russia\\
Institute of Physics,
Carl von Ossietzky University Oldenburg, Germany
}
\begin{abstract}
The index theorem implies that there are fermionic states localized on a soliton.
Presence of these modes may significantly alter the pattern of interaction between the solitons.
As a particular example we investigate the chiral magnetic Skyrmions coupled to spin-isospin fermions.
It is shown that there are sequences of fermionic modes localized on the Skyrmions.
We investigate the pattern of interaction between the soltions with localized modes and proved the existence of
stable system of magnetic Skyrmions bounded by the strong attractive dipole interaction
mediated by the chargeless fermionic modes.

\end{abstract}
\maketitle

{\it Introduction.~~}
The Skyrme model in 3+1 dimensions \cite{Skyrme:1961vq} is a prototype example of a field theory which supports
topological solitons, the Skyrmions. It was first proposed around 1960s to describe the strong
interactions of hadrons, in this framework the Skyrmions are identified with nucleons.
Later, it has been shown by Witten \cite{Witten:1983tw} that the Skyrme model
can be considered as a low energy effective theory of QCD in the limit of a large number of quark colours.
Such an effective theory can be constructed after integration over the
fermionic degrees of freedom, see e.g. \cite{Diakonov:1987ty},
the quarks do not appear as fundamental physical fields. In this paradigm
the contribution of the sea quarks to the baryon energy can be determined by considering the
normalizable bounded fermionic mode \cite{Kahana:1984dx,Kahana:1984be,Ripka:1985am},
which, according to the index theorem, always exist in the spectrum of fermions coupled to
the Skyrmion.

The pioneering work by Skyrme has opened new avenues for many areas of physics, similar
soliton configurations exist in various non-linear physical systems, see e.g. \cite{Manton:2004tk,Shnir2018}.
In particular, there is a simplified 2+1 dimensional analogue of the
original Skyrme theory, so called baby Skyrme model \cite{BB,Leese:1989gj,Bsk}.
This model finds various physical realizations, for example, planar Skyrmions occur in description of the
quantum Hall effect \cite{Hall,Lee:1990td}, such a  model also
arise in description of ferromagnetic structures \cite{Bogdanov}, or
in chiral nematic and anisotropic fluids \cite{Smalukh1,Smalukh2}.
Very recently, there has been rapidly increasing interest both in theoretical and experimental study of magnetic Skyrmions,
because of their possible use as information carriers in future magnetic storage devices, see \cite{Nago}.
In such a context the baby Skyrme model is no
longer considered as an effective low energy theory, constructed via integration over the fermionic degrees of freedom.
Further, general topological arguments based on the index theorem are given to support existence of quasi-zero
fermionic modes localized on the soliton (see e.g. \cite{Witten:1982fp,Novikov:1984ac}).
One can expect  the localized
fermionic modes may exist on magnetic Skyrmion \cite{Bogdanov} which has been experimentally
observed in chiral magnets with
Dzyaloshinskii-Moriya (DM) interaction \cite{skexp}.

Since the magnetic chiral Skyrmions always repel each other, there is no multisoliton solution in the model with
DM interaction term. However, an additional interaction
mediated by the fermions, may balance the scalar repulsion.

The Dirac electrons coupled to the planar magnetic Skyrmions were qualitatively discussed recently in
\cite{Hurst:2014tza,Andrikop}.
Such a system is of particular interest because it was argued that
localization of the fermions on Skyrmions may lead to a new mechanism of  non-conventional superconductivity in two-dimensional
systems \cite{Baskaran:2011wg,Lu,Khi}.
However, the symmetries of the DM interaction term
must be properly taken into account in investigation of the fermion - magnetic Skyrmion system.
A detailed analysis of the problem, is yet missing. It remains a major challenge
to construct complete set of solutions of the corresponding full system of dynamical equations,
especially for multisoliton configurations, which do not possess rotational invariance.

A main objective of this Letter  is to examine this system consistently, taking into account the
backreaction of the fermions.  In our numerical simulations we find the
spectrum of the corresponding Hamiltonian and
show that, indeed, there are various spin-isospin fermionic modes localized on a chiral magnetic Skyrmion.
Here we show, for the first time to our knowledge, that the these modes give rise to
additional attractive interaction between the
magnetic Skyrmions, which form bounded states. Notably, we do not find localized solutions for the chargeless
Dirac fermions coupled to the DM Skyrmions in two spatial dimensions.

{\it The model.~}
We consider the Hamiltonian, which describes a chiral magnetic planar system
with a violation of the inversion symmetry and a strong spin-orbit coupling of the compound:
\be
\mathcal{H}_s=\frac{J}{2}\left(\nabla\bfph\right)^2+D\bfph\cdot\left(\nabla\times\bfph\right)-\bf{B}\cdot\bfph \, .
\ee
These three terms correspond to the Heisenberg interaction, the DM interaction energy and
the symmetry breaking Zeeman energy of the interaction with an external magnetic field ${\bf B} = B \hat {\bf z}$, respectively.
Here $J$ is the magnetic stiffness constant and $D$ is the strength of the DM interaction. The magnetization vector
$\bfph$ is constrained to the surface of a sphere of unit radius: $\bfph \cdot \bfph=1$.
Since on the boundary the magnetization vector
is directed along the external magnetic field, ${{\bfph}}_\infty = (0,0,1)$, the field $\bfph$ is the
map $S^2$ to $S^2$. The corresponding topological invariant is
$Q= - \frac{1}{4\pi}\int \bfph \cdot   (\partial_x \bfph \times \partial_y \bfph) ~dxdy$, see e.g. \cite{Manton:2004tk,Shnir2018}.

The Lagrangian for the fermions, which are coupled to the magnetic Skyrmions, is
\be
\label{ferham}
\mathcal{H}_f=\Psi^\dagger\hat{\gamma}^3\left(-i\hat{\gamma}^k\partial_k+e\hat{\gamma}^k A_k+m+g\pmb{\tau}\cdot\bfph\right)\Psi.
\ee
Note that the fermion field $\Psi$ is a spin-isospin spinor, the isospin  matrices are defined as
$\pmb{\tau}=\mathbb{I}\otimes\pmb{\sigma}$, whereas the spin matrices are
$\hat{\gamma_\mu}=\gamma_\mu\otimes\mathbb{I}$. Here $\mathbb{I}$ is two-dimensional
identity matrix, $\slashed{\partial}=\hat{\gamma_\mu}\partial^\mu$ and $\pmb{\sigma}$ are
the usual Pauli matrices.  The last term in \re{ferham} represents the Hund coupling, the constant $g$
parameterizes the strength of the fermion-Skyrmion interaction,
$m$ is the fermion mass and $e$ is the electromagnetic coupling.
The vector potential of the external magnetic field is
$
A_\mu=\frac{B}{2}\left(0,-y,x\right)
$.
The total Hamiltonian of the coupled system $H=H_s+H_f$ depends on six parameters,
$J, D, B, g, m$ and $e$. An appropriate rescaling of the coordinates, fields and coupling constants
allows us to reduce the number of independent parameters to three: $g, m$ and $e$.

Hereafter we consider stationary configurations, $\bfph=\bfph(x^k),
\Psi=\psi(x^k)e^{-i\varepsilon t}$. The corresponding equation for the fermionic eigenfunctions is
\be
\label{Dirac}
\hat{\gamma}^3\left(-i\hat{\gamma}^k\partial_k+e\hat{\gamma}^k A_k+m+g\pmb{\tau}\cdot\bfph\right)\psi=\varepsilon\psi \, ,
\ee
thus, the eigenvalues $\varepsilon$ correspond to the energy of the fermions. Further, the equation for the field $\bfph$ is
\be
\label{spineq}
\Delta\bfph-2\nabla\times\bfph+\bfph_\infty-g\psi^\dagger\hat{\gamma_3}\pmb{\tau}\psi=0\, .
\ee
Note that
we neither impose the usual assumption that $\bfph$  is a fixed static background field, nor make
an approximation of its profile.

The complete system of coupled equations \re{Dirac},\re{spineq} must be solved numerically. This task can be simplified if we
take into account the symmetry properties of this system. First, we notice that the DM interaction
breaks the spatial and internal $\mathrm{O}(2)$ symmetries of the non-linear $\sigma$ model to the diagonal
subgroup. Secondly, the system enjoys the following discrete symmetries
\be
\begin{split}
x\to -x, &~\phi_y\to -\phi_y\, , \psi\to \hat{\gamma}^3\psi^{\ast}\, ,~
    \varepsilon\to -\varepsilon;\\
y\to -y, &~\phi_x\to -\phi_x\, ,\psi\to \tau_3\psi^{\ast}\, ,~ \varepsilon\to -\varepsilon;
\end{split}
\label{symm}
\ee
In the limiting case of massless uncharged fermions with $m=e=0$  the system \re{Dirac},\re{spineq} is enhanced by additional
symmetry of the fermion field $\psi\to -i\sigma_2\otimes\sigma_2\psi^\ast\, ,~ \varepsilon\to -\varepsilon$.

First, let us consider $\mathrm{O}(2)$ invariant configuration, which is parameterized by the ansatz:
\be
\label{ansSkgen}
\bfph=\left(\sin f(r) \cos\left( n\varphi+\delta\right), \sin f(r) \sin \left( n\varphi+\delta\right), \cos f(r)\right) \, .
\ee
Here $f(r)$ is some monotonically decreasing
radial function, $\varphi$ is the usual azimuthal angle, $n\in \mathbb{Z}$ and the phase
$\delta$ corresponds to the internal orientation of
the Skyrmion. Notably, the energy of the magnetic Skyrmion depends on $\delta$, it is minimal for $\delta=\pi/2$,
further, rotationally invariant configuration \re{ansSkgen} exists only for $n=1$ (helical Bloch Skyrmions \cite{Nago,magSkbook}).
Since the field must approach the vacuum on the spatial
asymptotic, it satisfies the boundary condition $\cos f(r) \to 1$ as $r \to \infty$, i.e., $f(\infty) \to 0$.


The corresponding rotationally invariant spin-isospin eigenfunctions with the eigenvalues $\varepsilon$
can be written as
\be
\label{ansFer}
\psi= \mathcal{N} \left(
\begin{array}{c}
v_1 e^{i l \varphi}\\
iv_2 e^{i (l+n) \varphi}\\
u_1 e^{i (l+1) \varphi}\\
iu_2 e^{i (l+n+1) \varphi}
\end{array}
\right)\, ,
\ee
where the components $u_i$ and $v_i$ are
functions of the radial coordinate only, $l\in\mathbb{Z}$ is the angular momentum of
fermion and $\mathcal{N}$ is a normalization factor, which is defined
from the usual condition
$
\int d^2x ~{\psi}^\dagger \psi
=2\pi {\mathcal{N}}^2 \int_0^\infty rdr (v_1^2+v_2^2+u_1^2+u_2^2)=1
$.

The rotationally invariant fermionic Hamiltonian \re{ferham} commutes
with the total angular momentum operator
$
K_3=-i\frac{\partial}{\partial\varphi}+\frac{\hat{\gamma_3}}{2}+\frac{\tau_3}{2}
$.
The corresponding half-integer eigenvalues $\kappa=1+l$ can be used to
classify different field configurations. The ground state corresponds to $\kappa=0$ and thus $l=-1$.

First, we note that there are localized modes among fermionic eigenfunctions \re{ansFer}.
Indeed, substitution of the ansatz \re{ansSkgen} into the equation  \re{spineq} yields
\begin{widetext}
\be
\label{spineqans}
f''+\frac{f}{r}-\sin f\left(B + \frac{\cos f-2r\sin f}{r^2}\right)+g\sin
f\left(u_1^2+v_2^2-u_2^2-v_1^2\right)+2g\cos f\left(v_1 v_2-u_1 u_2\right)=0 \, .
\ee
\end{widetext}
Linearizing this equation in the asymptotic region, where both $f$ and the
fermionic field profile functions approach zero, we obtain
\be
\label{spineqlin}
f''+\frac{f'}{r}-f\left(B + \frac{1}{r^2}\right)=0,
\ee
which is the usual modified Bessel equation, whose solution can be written in terms of the McDonald function, $f\sim K_1(r)$.
Thus,  as $r\to \infty$,  $f\sim\frac{e^{-\sqrt{B}r}}{\sqrt{r}}$, the Skyrmion is exponentially localized
for any form of the fermionic field and the asymptotic field of the magnetic Skyrmion may be thought of as generated by a
pair of orthogonal dipoles.

Considering fermions, we notice that in the limit of vanishing Hund
coupling the system is reduced to the usual Dirac fermions in the uniform magnetic field.
The energy spectrum of the charged fermions is given by the Landau levels
\be
\label{Landau}
\varepsilon_{k}^2=M^2+B\left(|e|(2k+1)\pm e\right),
\ee
where $k\in \mathbb{Z}$. The levels are twofold degenerated (except $k=0$), we can expect this degeneration is lifted for
non vanishing fermion-Skyrmion coupling.
The charged modes then become exponentially localized on the Skyrmion, this effect gives rise to
the electric charge of the configuration.

For the chargeless modes ($e=0$) the situation is different, this case is similar to that of the usual
fermion-Skyrmion system considered in \cite{Perapechka:2018yux}.
The asymptotic expansion of the equations
\re{Dirac} with parametrization \re{ansFer} at $r\to \infty$ yields
\be
\label{fermlin}
\begin{split}
    u_1 &\sim K_l(\sqrt{(g+m)^2- \varepsilon^2}\, r),\\
    u_2 &\sim K_{l+1}(\sqrt{(g-m)^2- \varepsilon^2}\, r),\\
    v_1 &\sim K_{l+1}(\sqrt{(g+m)^2- \varepsilon^2}\, r),\\
    v_2 &\sim K_{l+2}(\sqrt{(g-m)^2- \varepsilon^2}\, r)\, ,
\end{split}
\ee
and the exponentially localized modes exist as $|\varepsilon|<g-m$.

{\it Numerical results.~~}
The equations \re{spineq} and \re{Dirac}, together with constraint imposed by the normalization condition,
yield a system of integro-differential equations, which can be solved numerically. In a general case the system
is not rotationally invariant, then we impose an additional $\mathrm{O}(3)$ constraint on the scalar field $\bfph$.
We make use of the 4th order Newton-Raphson method implemented in the CESDSOL package, relative numerical errors  are
no higher than $10^{-5}$.

First, we consider rotationally-invariant system of  fermions coupled to the single Skyrmion.
\begin{figure}[t]
    \begin{center}
        \includegraphics[width=.235\textwidth, trim = 40 20 90 20, clip = true]{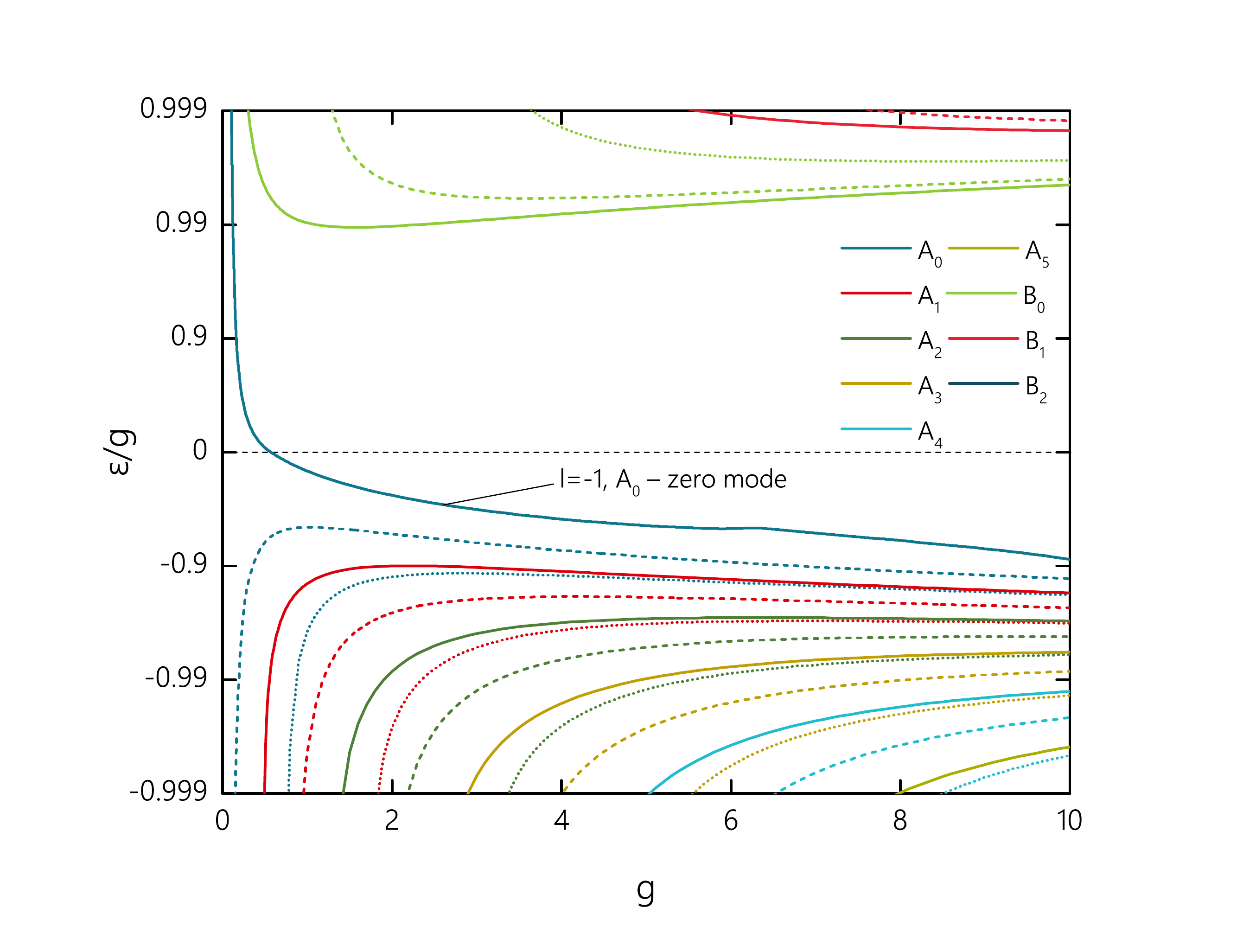}
        \includegraphics[width=.235\textwidth, trim = 40 20 90 20, clip = true]{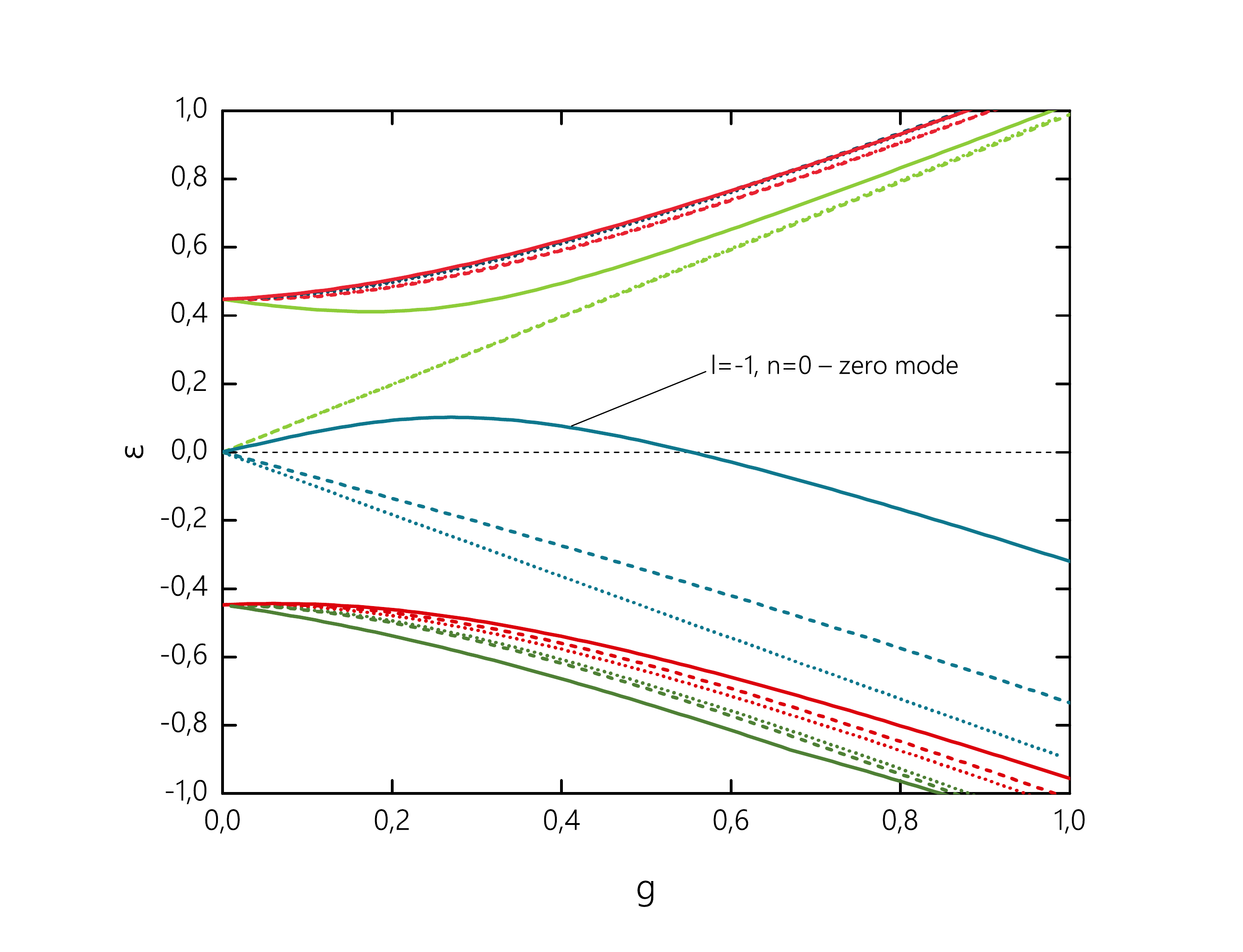}
    \end{center}
    \caption{\small
        Energy $\varepsilon$ of the chargeless ($e=0$, left plot)
        and charged ($e=-1$, right plot) localized massless fermions  as a function of the coupling constant
        $g$. The solid, dashed and dotted  lines correspond to $l=-1,0,1$, respectively. }
    \lbfig{Singlemodes}
\end{figure}

The results for the fermionic energy spectrum are presented in Fig.~\ref{Singlemodes}.
For the uncharged  modes the general pattern is similar to what we found in our previous study of the
fermions interacting with baby Skyrmions \cite{Perapechka:2018yux}. In agreement with the index theorem, for a given value
of $l$,
there is one zero-crossing mode which runs from positive to negative continuum as the Hund coupling $g$ is increasing.
Apart this mode there are localized states of two different types, which are linked to the negative and positive
continuum. We refer to them as the modes of the types $A$ and $B$, respectively.

The spectral flow of the charged
fermions is different, see Fig.~\ref{Singlemodes}, right plot.
In the limiting case $g=0$ the fermions are decoupled from the Skyrmion, they occupy the Landau levels \re{Landau}.
For each value of $l\neq 0$ there are two modes on each level, the zero mode corresponds to $k=0, l=-1$.
As the Hund coupling $g$ increases, the states start to deform, the energy of the lowest mode becomes positive,
it has a maximum at some value of $g$. As the coupling increases further, the energy of the lowest mode
is decreasing, it crosses zero at some critical value of the Hund coupling. This mode remains localized on the Skyrmion for
all values of the coupling, while other charged modes with $k\neq 0$ are linked to the positive or negative energy continuum,
approaching it at some set of critical values of the fermion-Skyrmion coupling $g$. Further, as $g$ increases,
the fermion-Skyrmion interaction becomes stronger than the interaction between the fermions and magnetic field ${\bf B}$,
thus there is a one-to-one correspondence between the corresponding localized modes and the uncharged modes of the types $A_k$ and $B_k$.

Localization of the fermionic modes may strongly affect the usual pattern of interaction between the magnetic Skyrmions.
Notably, even chargeless fermionic modes may balance the repulsive
interaction between the chiral Skyrmions.

Indeed, making use of the asymptotic equation \re{spineqlin} we can evaluate
the potential of the repulsive scalar interaction between the Skyrmions separated by the
distance $R$ as $V_{s}\sim K_0\left(\sqrt B R\right)$. Similarly, the asymptotic decay of the chargeless fermionic
field \re{fermlin} yields an additional channel of interaction. For solitons with two localized $A_0$-type modes the
corresponding potential is $V_{f}\sim\cos\theta K_0\left(\sqrt{(g+m)^2- \varepsilon^2} R\right)-K_0\left(\sqrt{(g-m)^2-
\varepsilon^2} R\right)$, where $\theta$ is an angle of relative orientations of the fermionic dipoles. Hence, this interaction
can be attractive and bounded system of solitons may exist for a certain set of values of the parameters of the system, for example
we can look for stable biskyrmion solutions, which represent
two chiral Skyrmions coupled by the localized fermionic modes.
\begin{figure}[t]
    \begin{center}
        \includegraphics[width=.23\textwidth, trim = 40 20 90 20, clip = true]{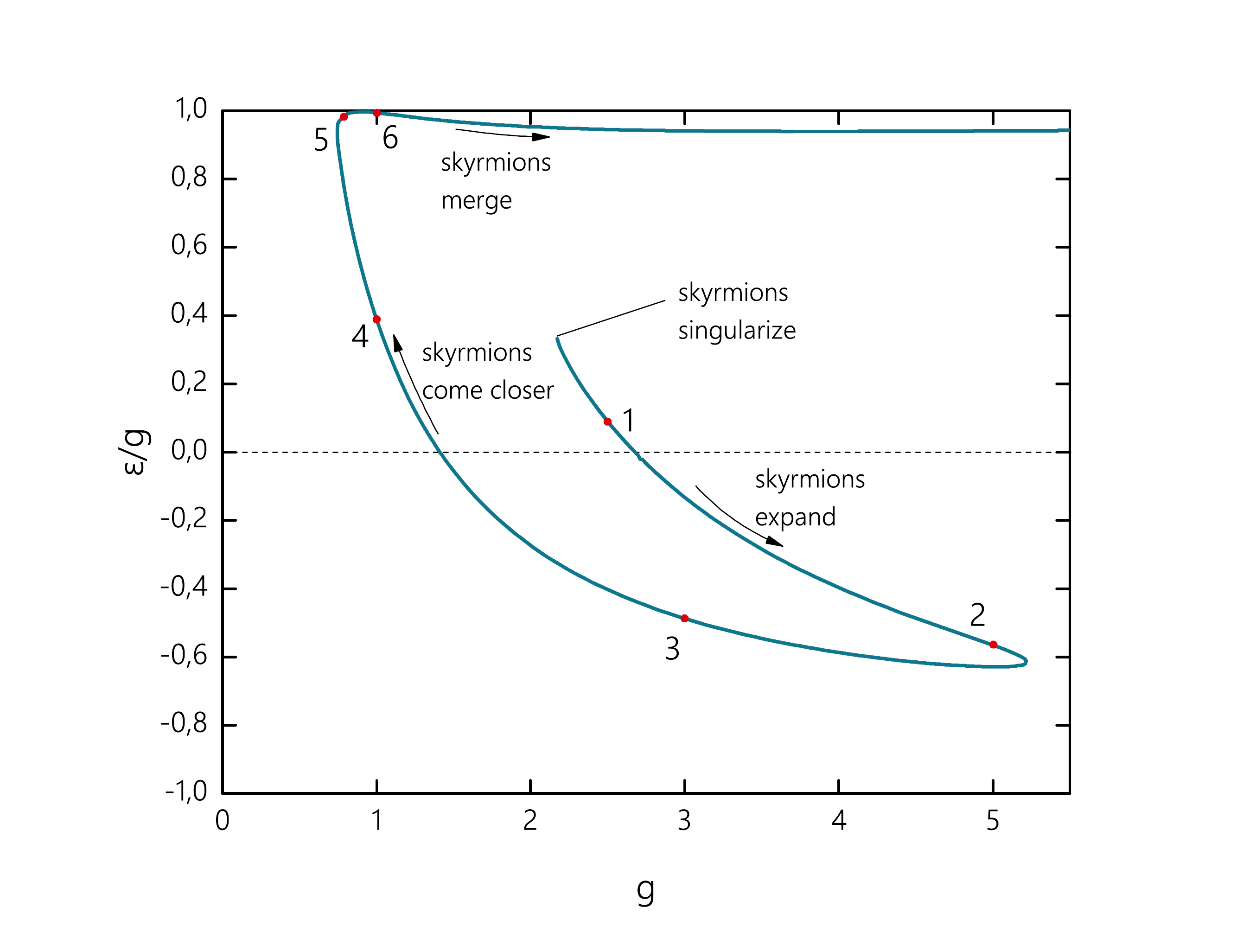}
        \includegraphics[width=.23\textwidth, trim = 40 20 90 20, clip = true]{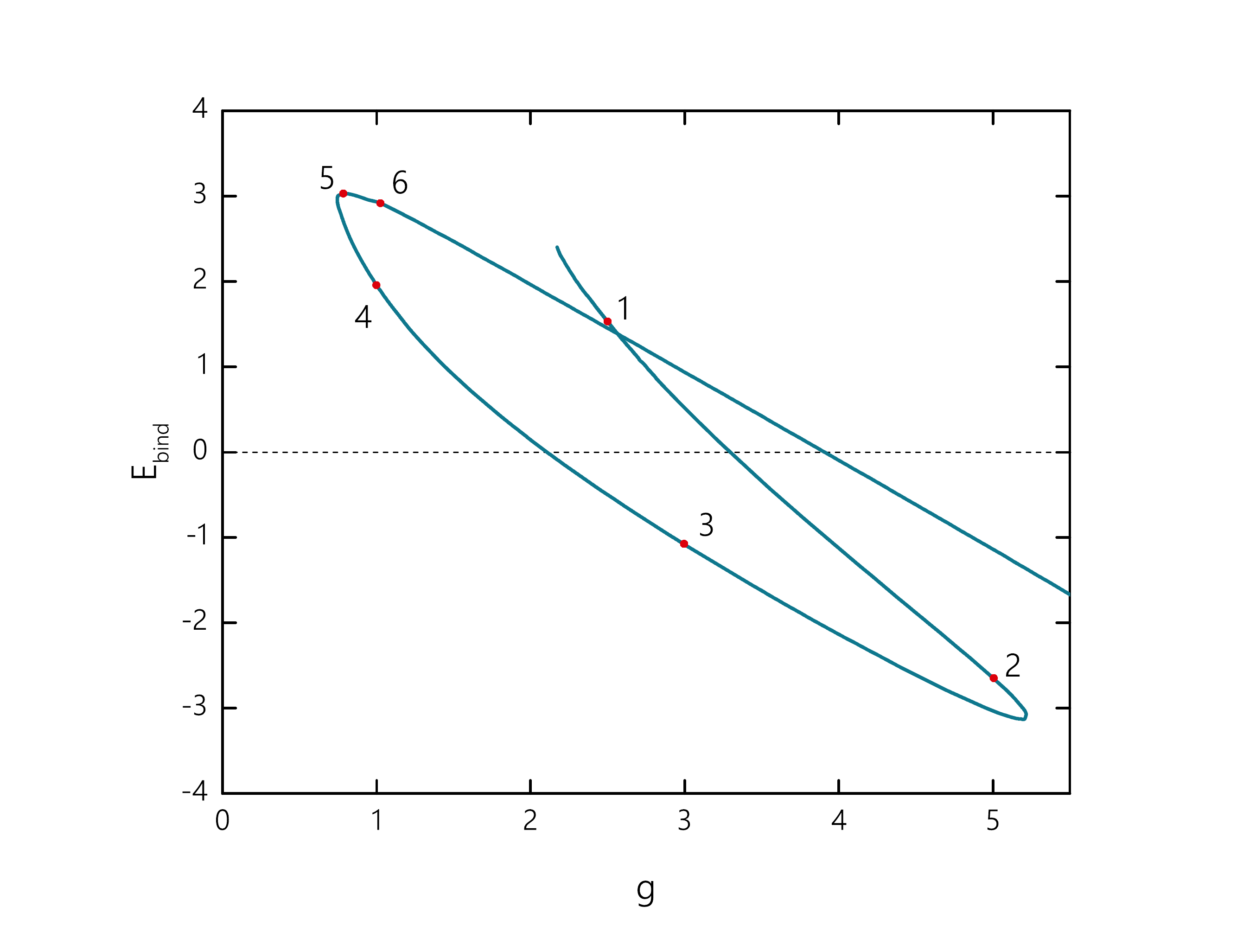}
    \end{center}
    \caption{\small
Normalized energy  ${\varepsilon}/{g}$ (left plot) of the chargeless massless fermionic mode coupled to a chiral biskyrmion
and the interaction energy $E_\mathrm{int}$ (right plot) of the solitons with localized mode
as a function of the coupling  $g$. The numbers on the curves correspond to the plots in Fig.~\ref{Bitop}}
    \lbfig{Bimodes}
\end{figure}

For the sake of simplicity we restrict our analysis to the massless uncharged fermions.
The energy of interaction of two Skyrmions in such a system  can be evaluated as
$E_\textrm{int}=E-2M-g$, where $E$ is the total energy of the biskyrmion-fermion configuration,
$M$ is the mass of a single Skyrmion and we take into account that for as $m=e=0$,
the lower bound of continuum is $\varepsilon_c=g$.

Our numerical investigation shows that biskyrmion-fermion configurations exist for relatively large values of
the coupling $g$,
moreover, there are several branches of solutions and different configurations may exist for the same value of $g$.
All solutions we found satisfy the symmetry restrictions \re{symm}.

In Fig.~\ref{Bimodes} we display the fermion energy (in unit of $g$) and
interaction energy $E_{int}$ as functions of the coupling strength $g$. First, we observe that
the bounded system of two charge one Skyrmions with two type $A_0$ modes localized on each of the solitons,
appears as a local minimum as $g$ increases above $g_{cr}^{(1)}=2.17$, see the left plots in Fig.~\ref{Bitop}.
The bounded solutions does not exist as $g <g_{cr}^{(1)}$.
As the Hund coupling increases, the overlap
between the modes becomes stronger and the solitons approach
each other. The energy of the fermionic modes is decreasing, it crosses zero and
the energy of interaction becomes negative although both Skyrmions remain separated, as shown in plots 2, Fig.~\ref{Bitop}. This
branch of solutions terminates
at some upper critical value of the coupling $g_{cr}^{(2)}\approx 5.22$, here it bifurcates with the second branch, which extends
backwards as the coupling $g$ is decreasing. Along this branch the maximum of the fermionic density distribution is located
at the center of the elongated soliton configuration, see plots 4 in Fig.~\ref{Bitop}.
The energy of the fermionic modes is increasing as $g$ is decreasing, in accordance with the index theorem it
crosses zero for a second time and tends towards the positive continuum as the solitons merge forming a rotationally
invariant configuration of topological degree 2, see plots 5, Fig.~\ref{Bitop}. At some lower critical value of the coupling
$g_{cr}^{(3)}\approx 0.75$ the symmetric combination of two fermionic modes approaches the continuum as it
bifurcates with the lowest rotationally invariant mode of the charge 2 configuration.
Along the corresponding third branch the coupling becomes stronger, the total energy of the configuration is decreasing.
No indication is found for termination of this branch, it exist for arbitrary large values of the coupling.

Furthermore, we found other multisoliton configurations bounded by the fermionic modes.
In Fig.~\ref{q3q4}, as particular examples, we display  the solutions we found in sectors of degrees $Q=3,4$, there is an
interesting pattern of transformations of the configurations as the Hund coupling varies.

In summary, we have shown that the localization of the spin-isospin fermionic modes on the magnetic
chiral Skyrmions with DM interaction strongly affects the usual picture of
interaction between the solitons.  We found multisoliton
solutions bounded by the localized fermionic modes and investigated the corresponding spectral flow. It must be stressed that
two-component fermions do not localize on the magnetic Skyrmion, spin-isospin symmetry plays a crucial role in appearance of
quasi-zero fermionic modes. Many issues remain for further study,
in particular, it is of great interest to construct magnetic Skyrmions lattice bounded by fermions.
Further, localization of the charged fermionic modes yields an electric charge of the soliton, thus
there is a long range Coulomb electric interaction between
two widely separated chiral Skyrmions with localized fermionic modes, which is supplemented by the short range scalar
repulsion. We hope to address this problems in our future
work. Finally, we expect similar bounded multisoliton solutions may exist in the conventional
baby-Skyrme model coupled with fermions,
as well as in various higher dimensional models, like for example in the Abelian Higgs model  .

{\it Acknowledgements.~~}
We are grateful to Steffen Krusch,  Nobuyuki Sawado and Paul Sutcliffe for helpful discussions.
Ya.S. gratefully acknowledges the support of the Alexander von Humboldt Foundation and from the
Ministry of Education and Science of Russian Federation, project No 3.1386.2017.

\newpage

\onecolumngrid

\begin{figure}
\hspace{0mm} 1 \hspace{27mm} 2\hspace{27mm} 3\hspace{27mm} 4\hspace{27mm} 5\hspace{27mm} 6
        \includegraphics[width=.16\textwidth]{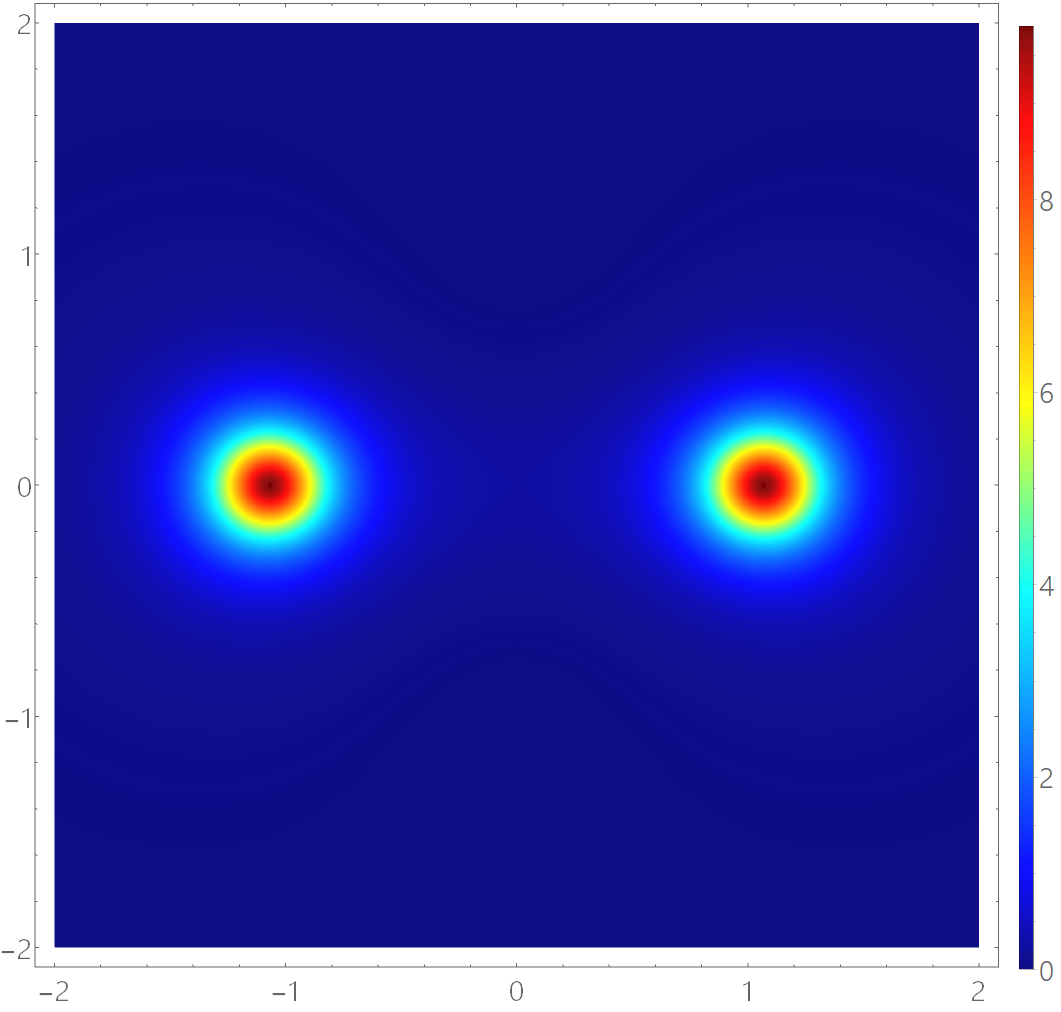}
        \includegraphics[width=.16\textwidth]{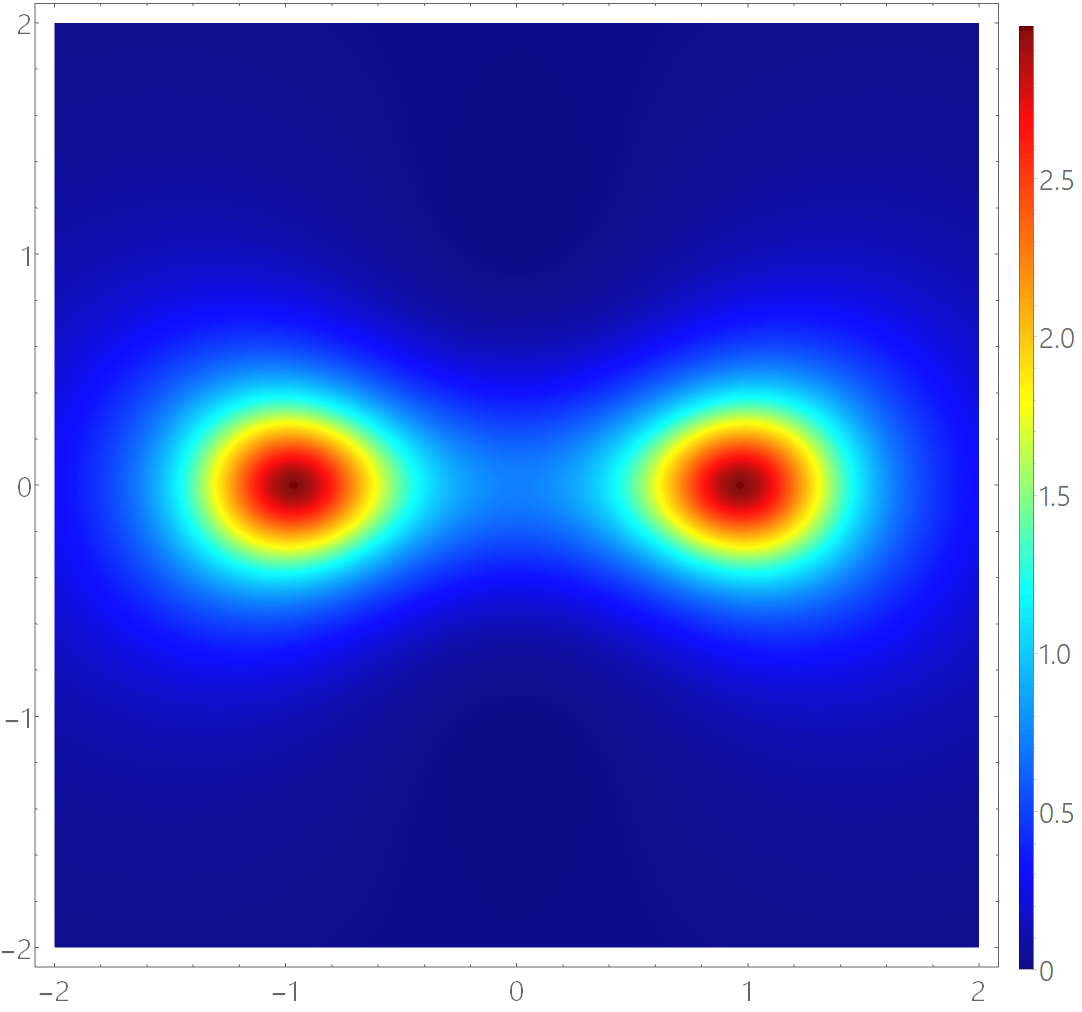}
        \includegraphics[width=.16\textwidth]{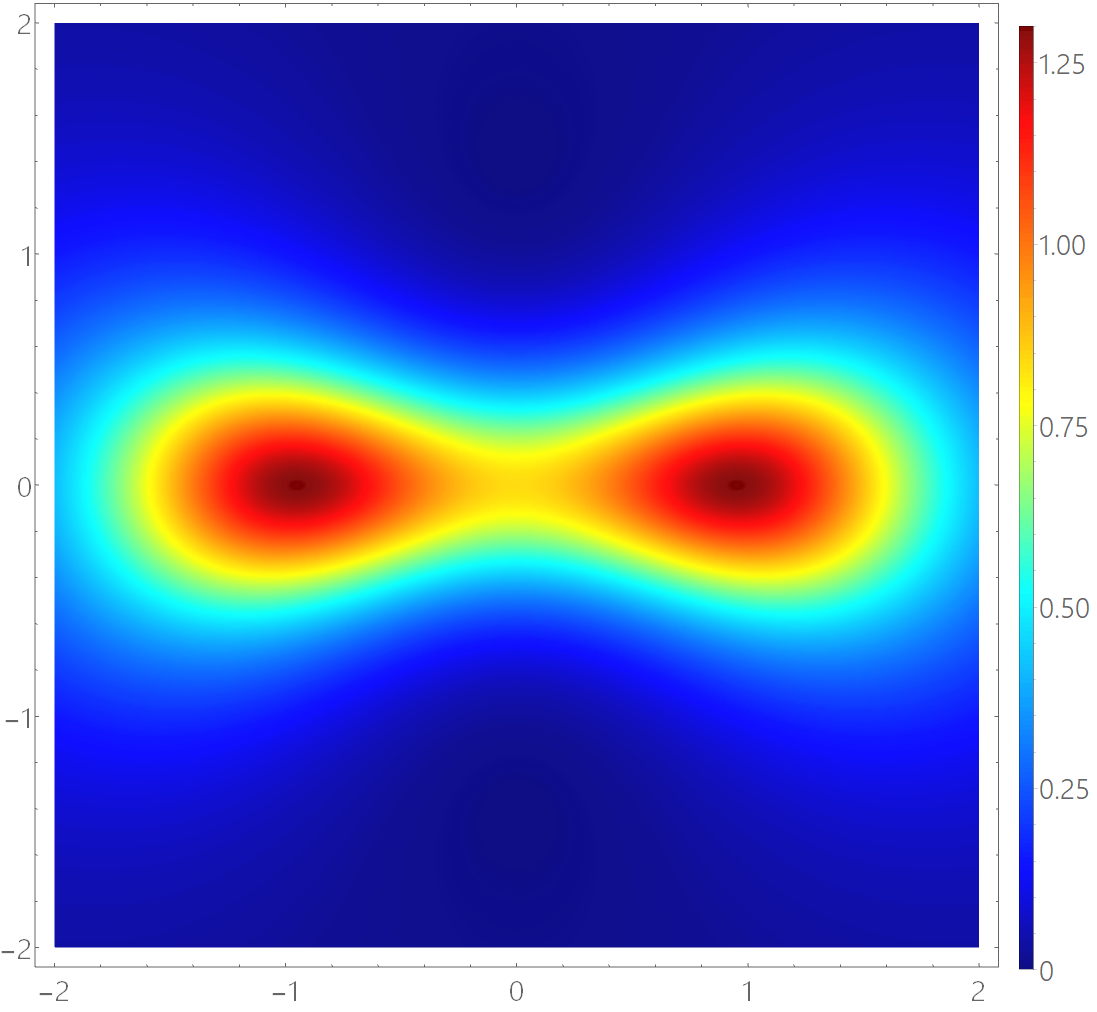}
        \includegraphics[width=.16\textwidth]{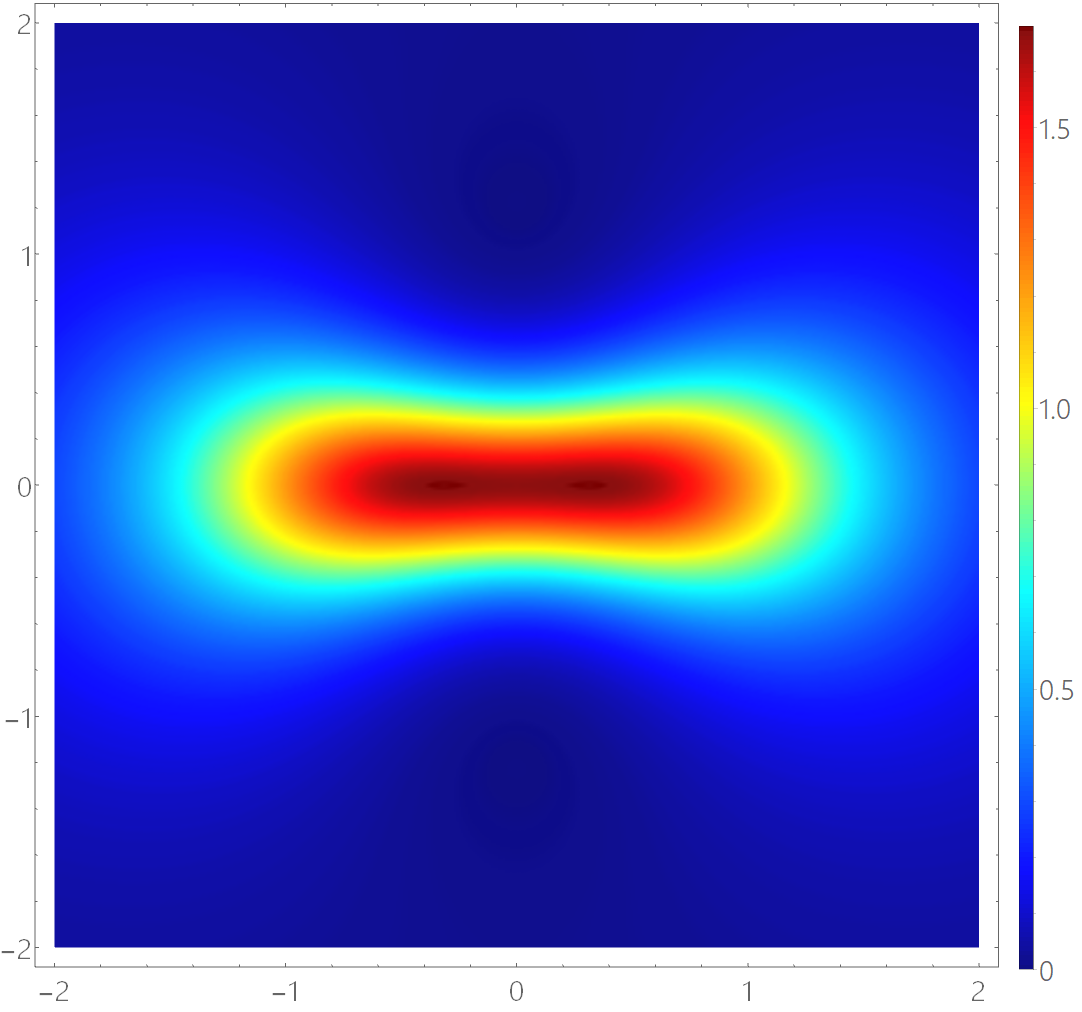}
        \includegraphics[width=.16\textwidth]{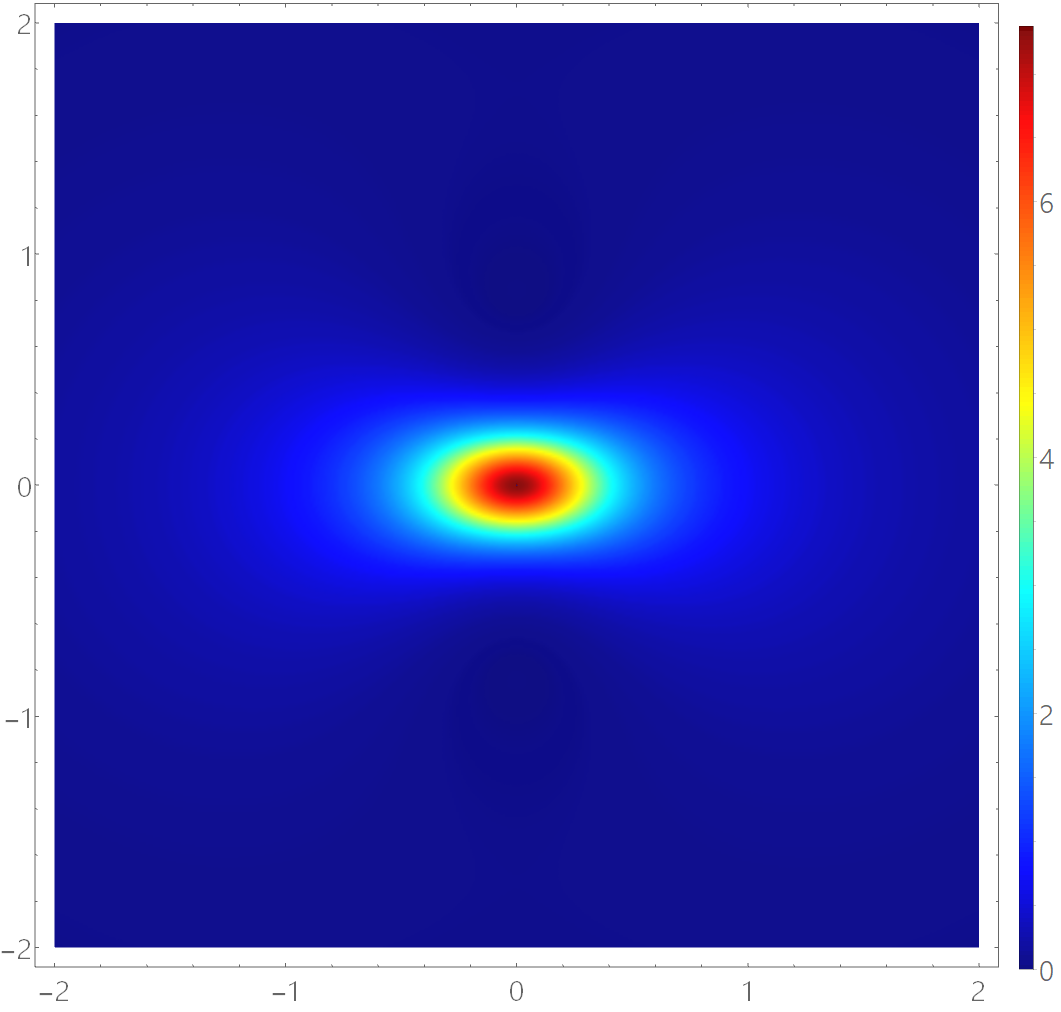}
        \includegraphics[width=.16\textwidth]{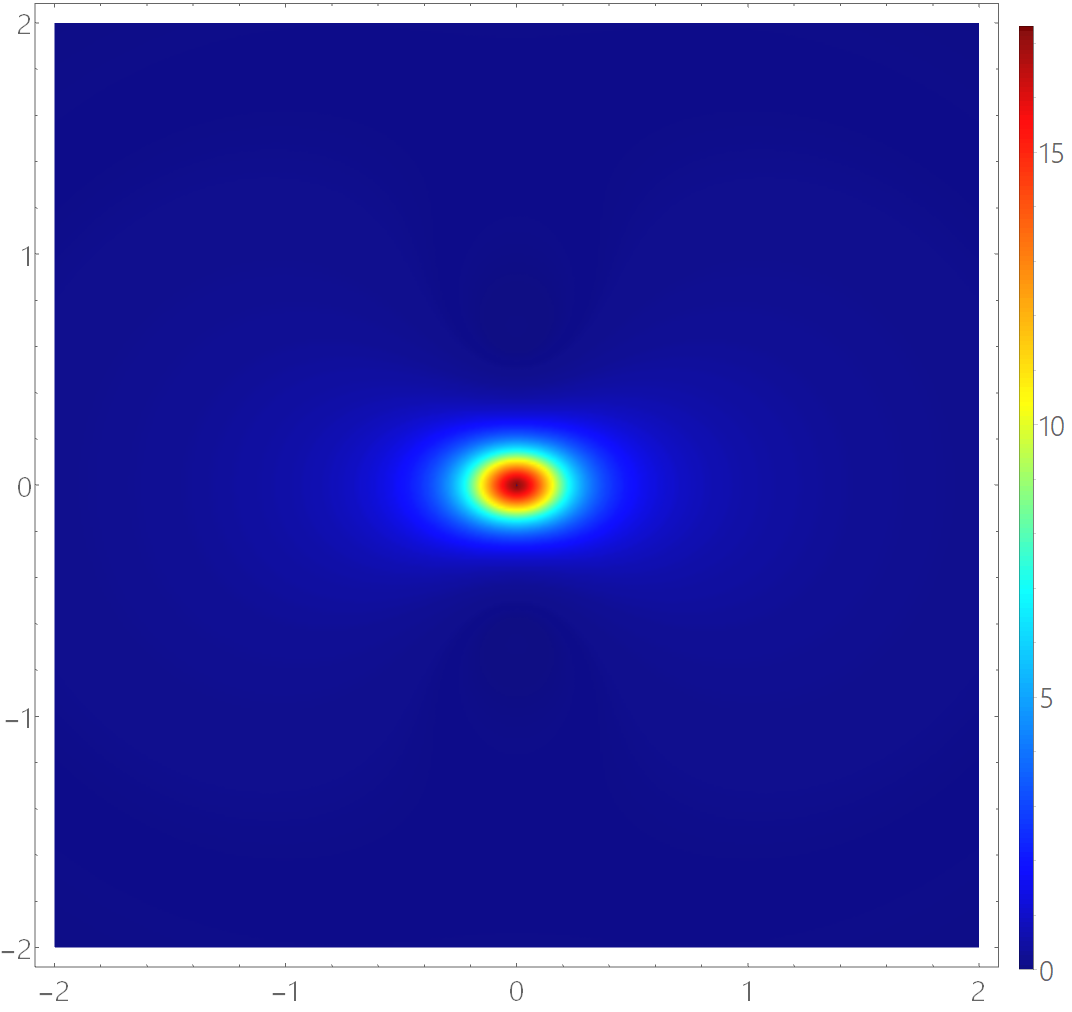}
        \includegraphics[width=.16\textwidth, clip = true]{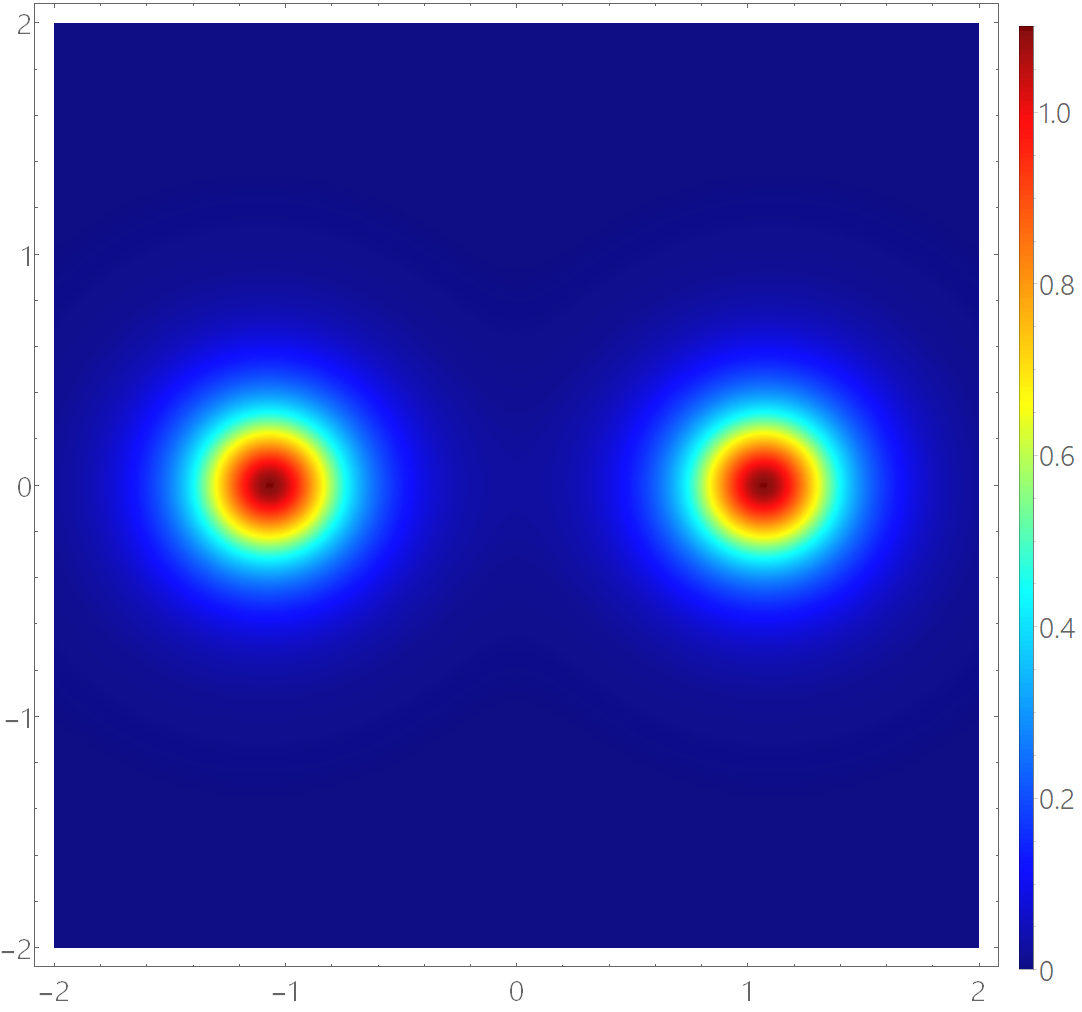}
        \includegraphics[width=.16\textwidth, clip = true]{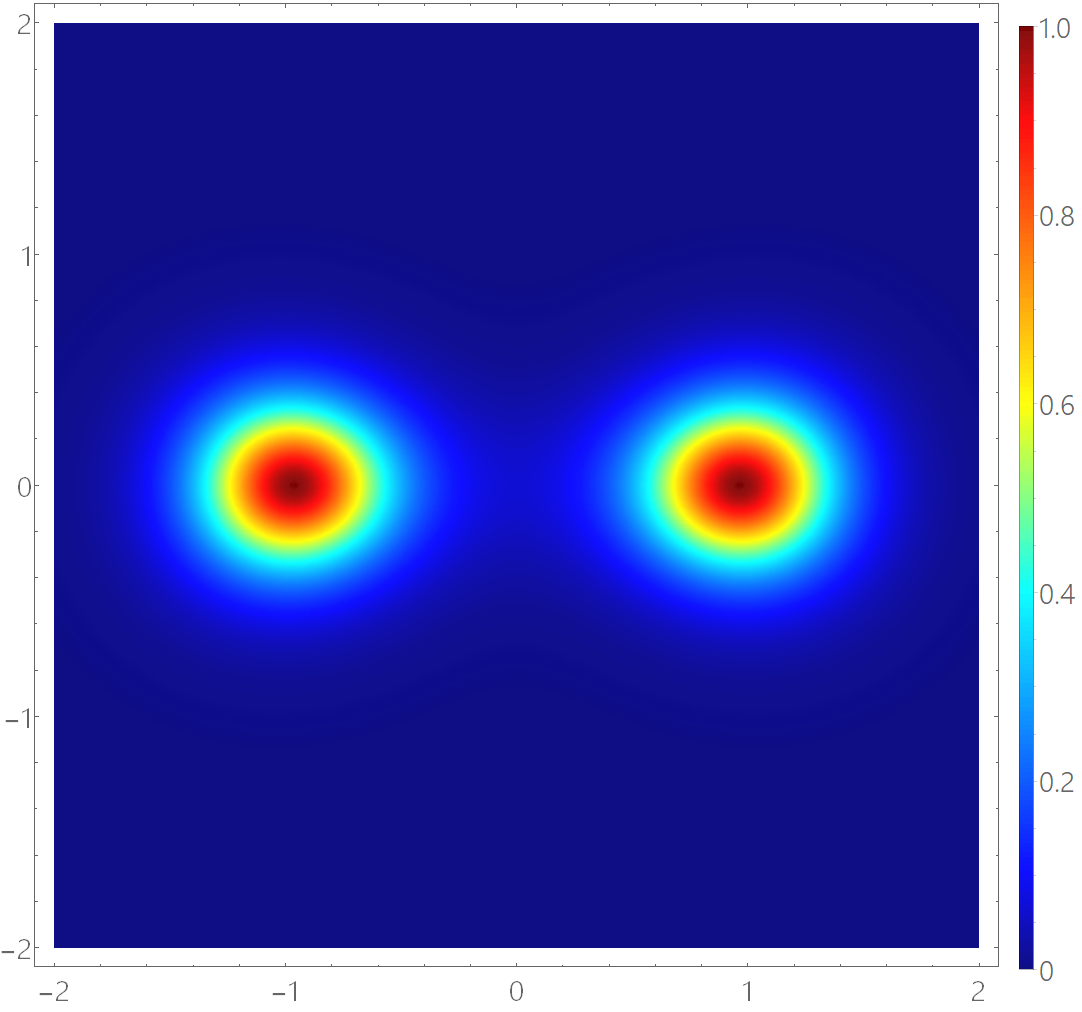}
        \includegraphics[width=.16\textwidth, clip = true]{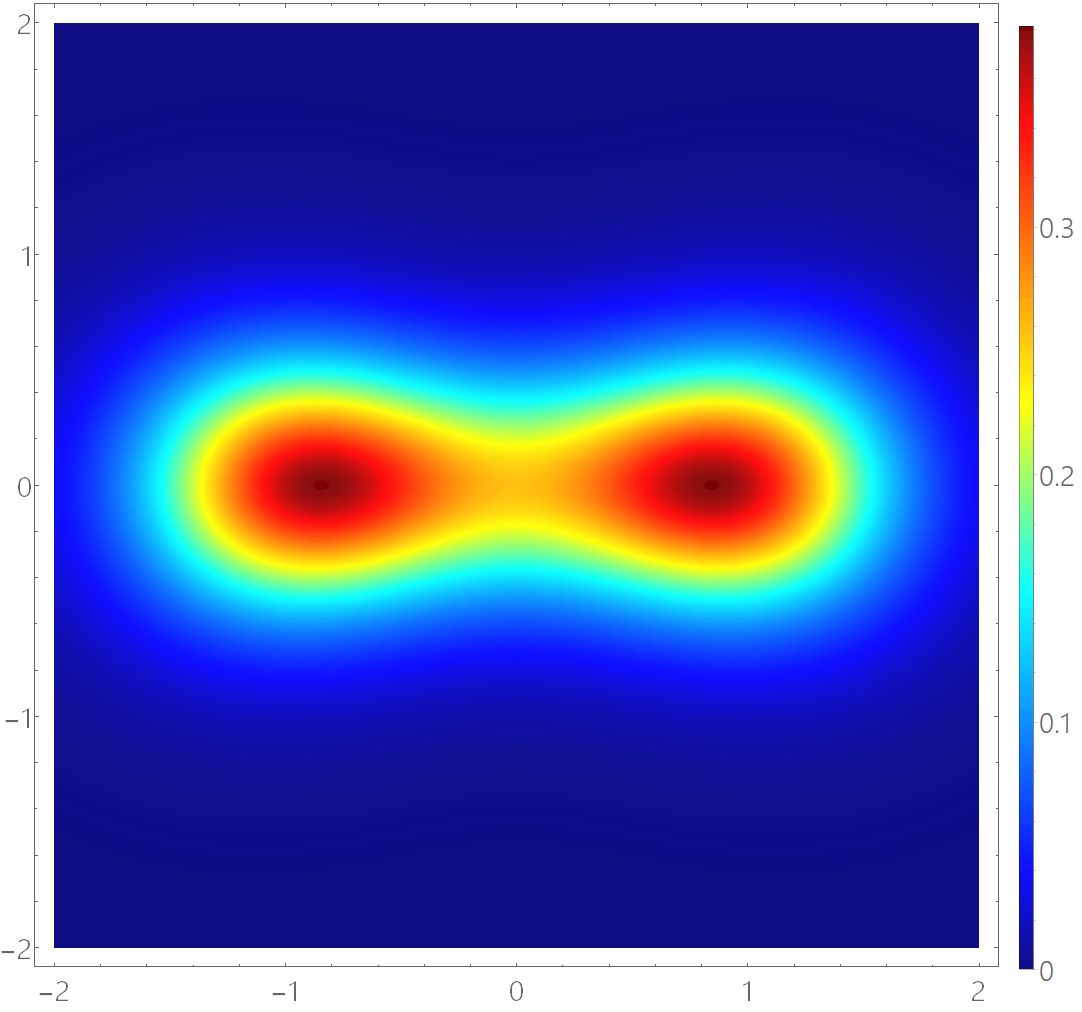}
        \includegraphics[width=.16\textwidth, clip = true]{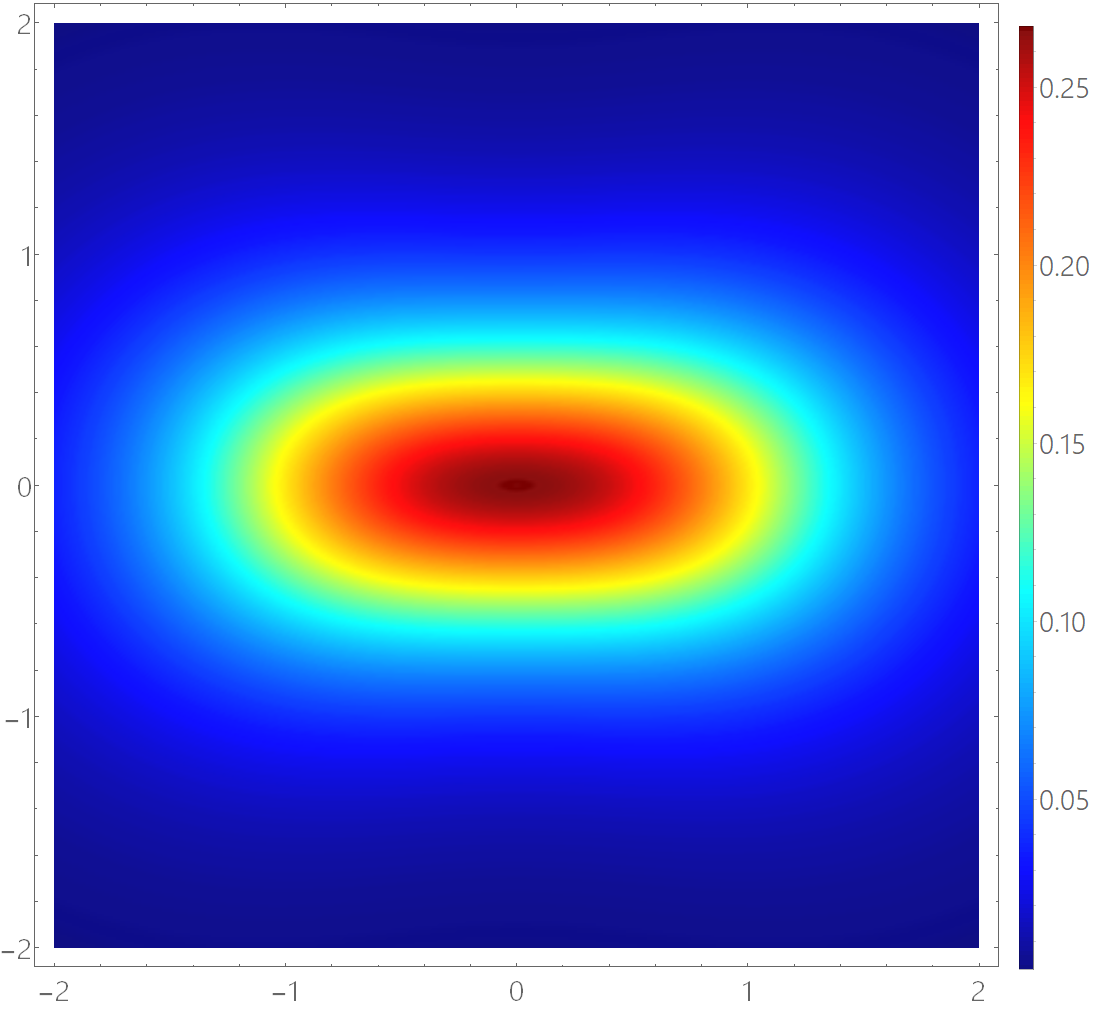}
        \includegraphics[width=.16\textwidth, clip = true]{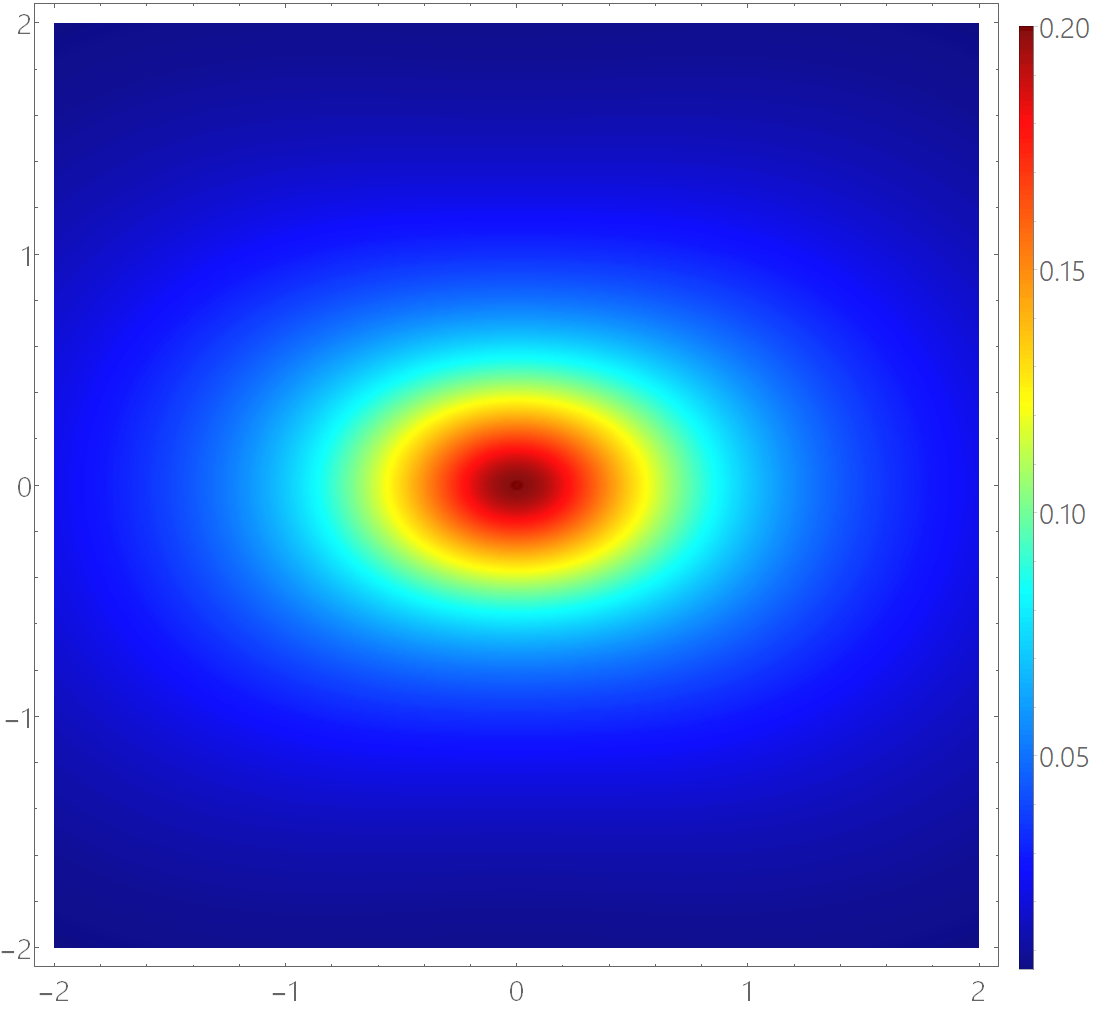}
        \includegraphics[width=.16\textwidth, clip = true]{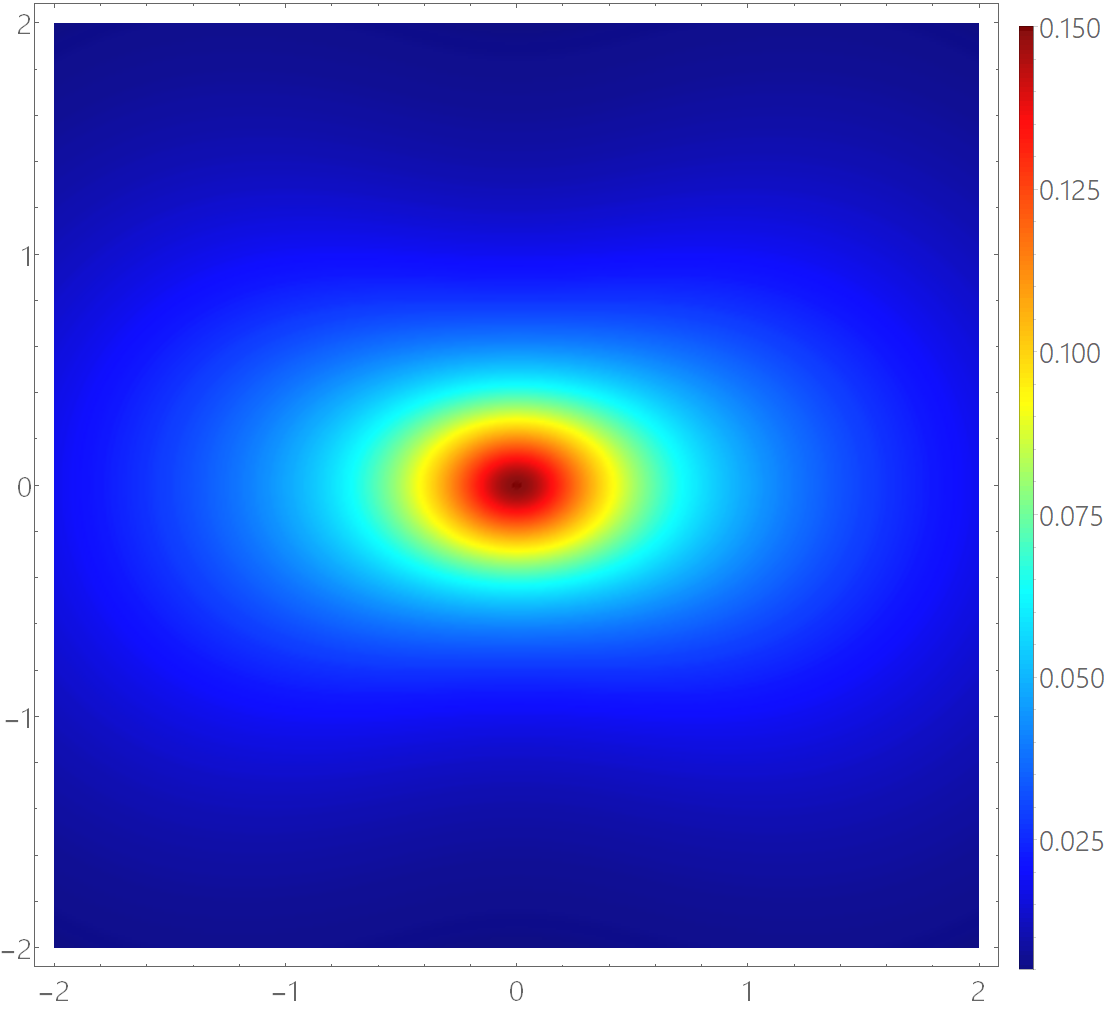}
    \caption{\small Biskyrmion-fermion bounded system:
        density plots of the distributions of the topological charge density  (upper row) and the
        fermionic density (bottom row). The plot numbers correspond to the
        values of the coupling constants marked as dots in Fig.~\ref{Bimodes}.}
    \lbfig{Bitop}
\end{figure}

\twocolumngrid

\begin{figure}[t]
    \begin{center}
        \includegraphics[width=.23\textwidth]{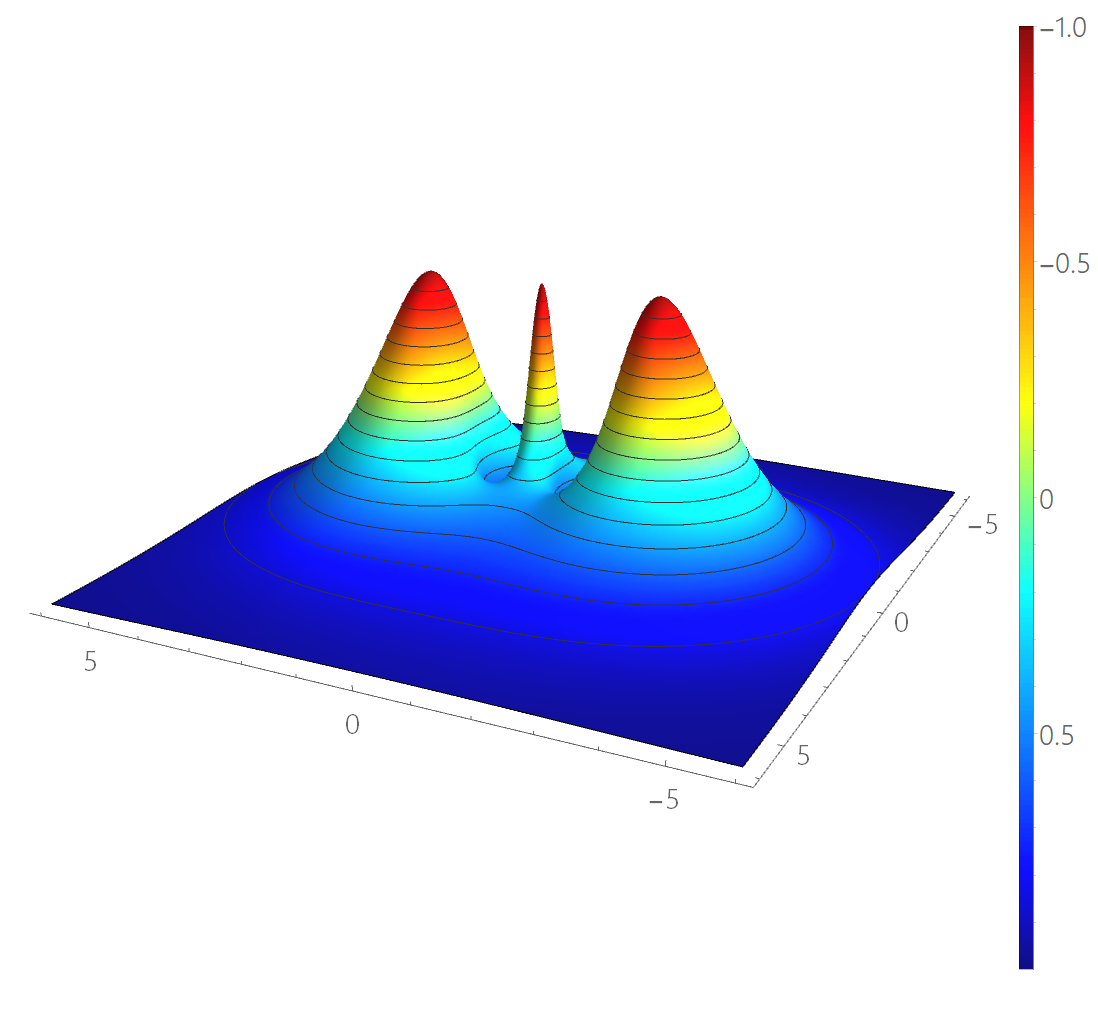}
        \includegraphics[width=.23\textwidth]{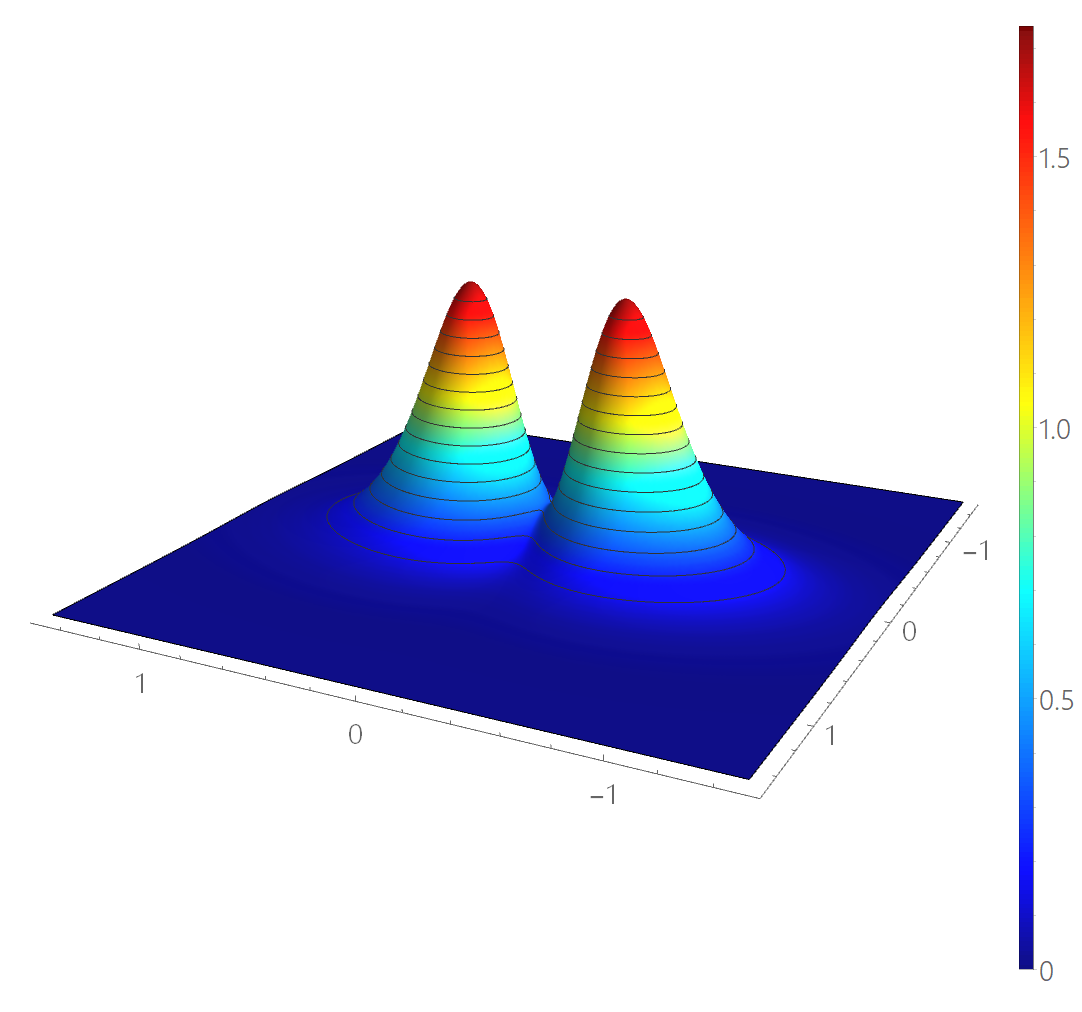}
        \includegraphics[width=.23\textwidth]{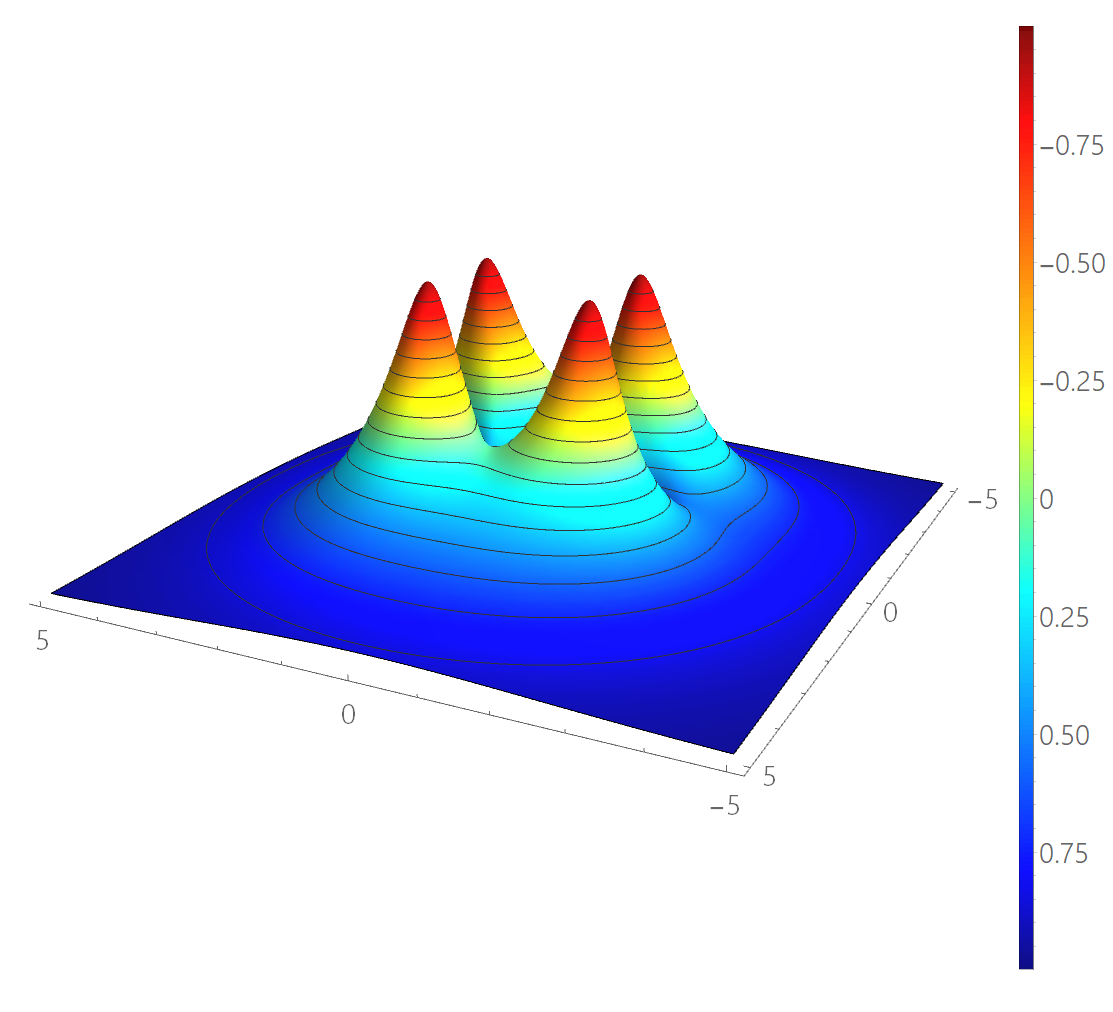}
        \includegraphics[width=.23\textwidth]{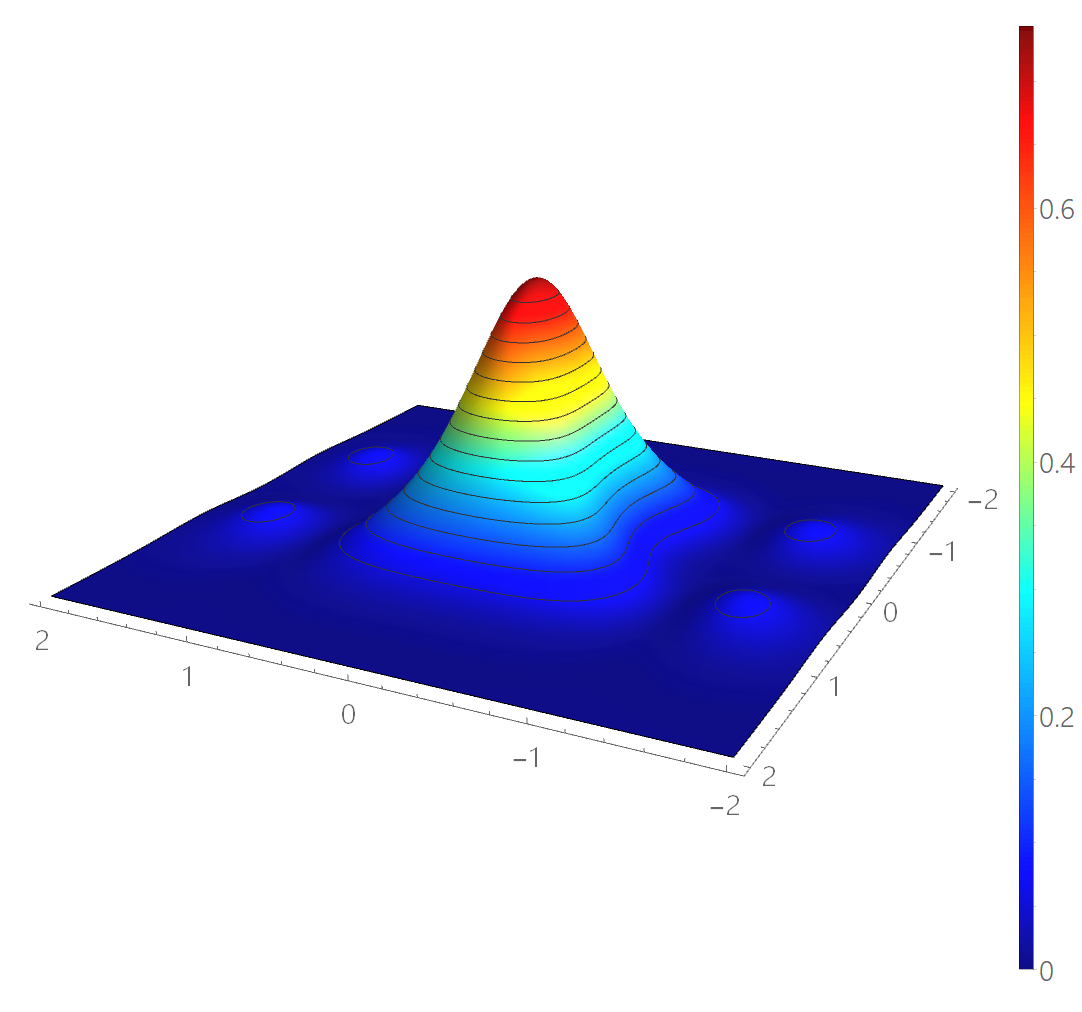}
    \end{center}
    \caption{\small
        $Q=3$ (upper row) and $Q=4$ (lower row) multiskyrmion-fermion bounded systems:
        Field components $\phi_3$ of the magnetic Skyrmions (left column) and the
        corresponding fermionic density distributions (right column) at $g=5$.}
    \lbfig{q3q4}
\end{figure}


\begin{thebibliography}{00}
\bibitem{Skyrme:1961vq}
  T.~H.~R.~Skyrme,
  Proc.\ Roy.\ Soc.\ Lond.\ A {\bf 260} (1961) 127.
\bibitem{Witten:1983tw}E.~Witten,
Nucl.\ Phys.\ B,  {\bf 223} (1983) 422
\bibitem{Diakonov:1987ty}
  D.~Diakonov, V.~Y.~Petrov and P.~V.~Pobylitsa,
  Nucl.\ Phys.\ B {\bf 306} (1988) 809
\bibitem{Kahana:1984dx}
S.~Kahana, G.~Ripka and V.~Soni,
  Nucl.\ Phys.\ A {\bf 415}  (1984) 351
\bibitem{Kahana:1984be}
  S.~Kahana and G.~Ripka,
  Nucl.\ Phys.\ A {\bf 429} (1984) 462
\bibitem{Ripka:1985am}
  G.~Ripka and S.~Kahana,
  Phys.\ Lett.\  {\bf 155B} (1985) 327
\bibitem{Manton:2004tk}
  N.~S.~Manton and P.~Sutcliffe,
  {\it 'Topological solitons',}
    Cambridge University Press, 2004.
\bibitem{Shnir2018}
Y.M.~Shnir,
{\it 'Topological and Non-Topological Solitons in Scalar Field Theories'},
Cambridge University Press, 2018.
\bibitem{BB}
A.A.~Bogolubskaya  and I.L.~Bogolubsky,
Phys.\ Lett.\ A  {\bf 136} (1989) 485
\bibitem{Leese:1989gj}
R.A.~Leese, M.~Peyrard and W.J.~Zakrzewski,
Nonlinearity {\bf 3} (1990) 773
\bibitem{Bsk}
B.~M.~A.~G. Piette, B.J.~Schroers and W.J.~Zakrzewski,
Z. Phys. C  {\bf 65} (1995) 165
\bibitem{Hall}A.~Schmeller, J.P.~Eisenstein,
L.N.~Pfeiffer, and K.W.~West
Phys.\ Rev.\ Lett.\   {\bf 75} (1995) 4290
\bibitem{Lee:1990td}
  D.~H.~Lee and C.~L.~Kane,
  Phys.\ Rev.\ Lett.\  {\bf 64} (1990) 1313
\bibitem{Bogdanov} A.N.~Bogdanov and D. A.~Yablonsky,
Sov.\ Phys.\ JETP  {\bf 95} (1989) 178  ;\\
A.N.~Bogdanov,
JETP Lett. {\bf 62} (1995) 247
\bibitem{Smalukh1}I.I.~Smalyukh et al,
Nat. Mater. {\bf 9} (2010) 139
\bibitem{Smalukh2}
P.J.~Ackerman et al,
Phys. Rev. E {\bf 90} (2014) 12505
\bibitem{Nago}N.~Nagaosa and Y.~Tokura,
Nature Nanotechnology {\bf 8} (2013) 899
\bibitem{Witten:1982fp}
  E.~Witten,
  Phys.\ Lett.\ B {\bf 117} (1982) 324
\bibitem{Novikov:1984ac}
  V.~A.~Novikov, M.~A.~Shifman, A.~I.~Vainshtein and V.~I.~Zakharov,
  Phys.\ Rept.\  {\bf 116} (1984) 103
\bibitem{Perapechka:2018yux}
  I.~Perapechka, N.~Sawado and Y.~Shnir,
  JHEP {\bf 1810} (2018) 081
\bibitem{skexp}S.~M\"{u}hlbauer et al,
Science  {\bf 323} (2009) 915;\\
X.Z.~Yu, Y.~Onose, N.~Kanazawa, J.H.~Park, J.H.~Han,
Y.~Matsui, N.~Nagaosa, and Y.~Tokura,
Nature {\bf 465} (2010) 901
\bibitem{Hurst:2014tza}
  H.~M.~Hurst, D.~K.~Efimkin, J.~Zang and V.~Galitski,
  Phys.\ Rev.\ B {\bf 91} (2015) no.6,  060401
\bibitem{Andrikop}D.~Andrikopoulos, B.~Sor\'{e}e, and J.~De Boeck,
Journal of Applied Physics, {\bf 119} (2016) 193903.
\bibitem{Baskaran:2011wg}
  G.~Baskaran,
arXiv:1108.3562.
\bibitem{Lu}C.K.~Lu  and I.F.~Herbut,
Phys.\ Rev.\ Lett.\   {\bf 108} (2012) 266402.
\bibitem{Khi}A.~Knigavko, B.~Rosenstein, and Y.F.~Chen,
Phys.\ Rev.\  B {\bf 60} (1999) 550
\bibitem{magSkbook}J.P.~Liu, Z.~Zhidong  and Z.~Guoping (eds).
\textit{Skyrmions: topological structures, properties, and applications},
CRC Press, 2016.
\end{thebibliography}
\end{document}